\documentclass[aps,prl,twocolumn,superscriptaddress,floatfix,citeautoscript,preprintnumbers]{revtex4-2}

\usepackage{graphicx,color}
\usepackage{amssymb}
\usepackage{amsmath}
\usepackage{epstopdf}
\usepackage[normalem]{ulem}
\usepackage{comment}
\usepackage{breakcites}
\usepackage[breaklinks=true]{hyperref}
\usepackage[caption=false]{subfig}
\usepackage{enumitem}
\usepackage{soul}
\usepackage{bbold,dsfont}
\usepackage{siunitx}
\usepackage{braket}

\begin{document}

\title{Local Certification of Many-Body Steady States}

\begin{abstract}

We present a relaxation-based method to bound expectation values on the steady state of dissipative many-body quantum systems described by master equations of the Lindblad form. Instead of targeting to represent the entire state, we promote the reduced density matrices to our variables and enforce the constraints that are imposed on them by consistency with a global steady state. The resulting constraints have the form of a semidefinite program, which allows us to efficiently bound the values a given expectation value can take in the steady state. Our results show fast convergence of the bounds with the size of the reduced density matrices, giving very competitive predictions for the steady state of several one- and two-dimensional models for an arbitrary number of particles.

\end{abstract}

\author{Miguel Frías Pérez}\email{miguel.frias@icfo.eu}
\affiliation{ICFO-Institut de Ciències Fotòniques, The Barcelona Institute of Science and Technology, 08860 Castelldefels (Barcelona), Spain}

\author{Antonio Ac\'{i}n}
\affiliation{ICFO-Institut de Ciències Fotòniques, The Barcelona Institute of Science and Technology, 08860 Castelldefels (Barcelona), Spain}
\affiliation{ICREA-Institució Catalana de Recerca i Estudis Avan\c cats, Lluís Companys 23, 08010 Barcelona, Spain}

\maketitle

\paragraph{Introduction.---} The description of interacting many-body quantum systems is one of the fundamental challenges of quantum mechanics, as it involves a Hilbert space whose dimension grows exponentially with the number of constituents. While this already renders the treatment of isolated systems difficult, the problem becomes particularly severe when the system under consideration is coupled to an environment~\cite{breuer2002, nielsen2010}. In this case, the description of the state of the system is given by the density matrix, which has a number of coefficients that scales quadratically with the dimension of the Hilbert space. Furthermore, these coefficients are coupled together by the requirement that the density matrix is positive, a nontrivial constraint that requires knowledge of the spectrum of the matrix. 

Despite the challenge, developing classical descriptions of open many-body quantum systems is of interest. On the one hand, such systems are relevant to the current experiments in quantum simulation and quantum computation, where large numbers of quantum degrees of freedom are engineered to interact in a controlled manner. Although these platforms aim to be as isolated as possible, residual decoherence and dissipation are unavoidable. This creates a need for tools to certify and benchmark the performance of quantum devices in realistic conditions. Furthermore, these classical tools help us understand the boundary between quantum and classical computational power. On the other hand, coupling to an environment can also be viewed as a resource. Appropriately engineered dissipative processes can be used to drive a system toward desired states~\cite{zhan2026, rouze2025}, to perform dissipative quantum computation~\cite{verstraete2009, diehl2008} or to create phases of matter without an equilibrium counterpart~\cite{sieberer2025}. Together, these two perspectives motivate the search for schemes to describe the non-equilibrium dynamics of open many-body quantum systems classically. 


This motivation has led to the development of many numerical approaches~\cite{weimer2021, fazio2025}. The exact description of the system is limited to small sizes, even smaller than what is achievable in the case of pure states. Tensor networks, which provide a compact representation of quantum states when the amount of entanglement in them is limited, have been proposed for this task~\cite{zwolak2004, verstraete2004, cui2015}. However, they suffer from fundamental limitations in the simulation of dynamics, are limited in higher-dimensional systems, and are not easily combined with positivity~\cite{osborne2006, schuch2008, kliesch2014, kilda2021}. Other methods, including neural network states~\cite{yoshioka2019, hartmann2019, vicentini2019, nagy2019} or phase-space methods~\cite{polkovnikov2010, schachenmayer2015, huber2021}, are regularly used. However, just as the rest of the methods presented, they are heuristic in nature and lack performance guarantees.


In recent times, the idea of using convex relaxations as a way to solve many-body quantum systems has gained traction. The insight behind them is to abandon a complete description of the quantum state and focus instead on their relevant correlations, e.g., low-order correlation functions or local marginals of the state. Clearly, the number of parameters required to specify them does not grow exponentially with the number of constituents. However, this information is still constrained by the algebra of the operators and by the positivity of the global state. Relaxation-based methods explore the different values this data can take, provided we still respect the underlying constraints. The potential advantage is that, contrary to variational methods, they provide rigorous results, in the form of bounds. Examples of this approach are~\cite{anderson1951, mazziotti2001, nakata2001, barthel2012, baumgratz2012, wang2024, kull2024, eisert2023, haim2020, han2020, rai2024, jansen2025, mortimer2026} for the ground-state problem, \cite{fawzi2024} for thermal states or \cite{lawrence2024, mishra2024, araujo2025} for the time-evolution of many-body quantum systems.


In this work, following \cite{mortimer2025, robichon2024, mok2025, lau2023, cho2025}, we present a relaxation-based scheme to compute local expectation values in the steady state of an open quantum system described by a Lindblad master equation. The construction can be applied to systems of arbitrary dimension and number of constituents. We benchmark it with several models in one- and two-dimensional systems, observing fast convergence of the bounds with the size of the reduced density matrices. In the next section, we present our notation and the setting for our technique.


\paragraph{Master equations and steady states.---}Our algorithm applies to systems whose dynamics are governed by a Markovian equation of the Lindblad form \cite{breuer2002}, $d\rho / dt = \mathcal{L}\left(\rho\right)$, where the right-hand side is the Lindbladian superoperator, 
\begin{equation}
\mathcal{L}\left( \rho \right) = -i\left[H, \rho \right] + \sum_n \mathcal{D}_n \left( \rho \right) .
\end{equation} The Hamiltonian $H = \sum_n h_n$ governs the coherent part of the dynamics, while the second term, $~{\mathcal{D}_n \left( \rho \right) = L_n \rho L^\dagger_n - \frac{1}{2} \{L^\dagger_n L_n, \rho \} }$, with $L_n$ the so-called jump operators, models the interaction between the system and its environment. We consider many-body systems where individual quantum systems of arbitrary dimension, also known as qudits, are arranged in a lattice and the terms in the Lindbladian are local operators, that is, they have support (act nontrivially) in a small number of contiguous sites. 

Steady states are fixed points of the evolution, $d\rho_s / dt = 0$, and thus correspond to positive operators that are in the kernel of the Lindbladian superoperator, 
\begin{equation}\label{eq:stst}
    \mathcal{L}\left(\rho_s\right) = 0.
\end{equation} In general, finding $\rho_s$ amounts to solving the previous linear equation. However, for many-body systems, the dimension of the operators and superoperators grows exponentially with the number of constituents of the system and this quickly becomes an impractical approach. In this work, instead of dealing with the exponential complexity of the density matrix of the steady state, we take a complementary route: we promote the reduced density matrices of the system to our variables and look for the constraints that are imposed on them by consistency with the global steady state, defined by Eq.~\eqref{eq:stst}. In what follows, we first present our method in a translation-invariant one-dimensional system. Afterwards, we will discuss how to generalize it to non-symmetric cases and higher dimensions.  

\begin{figure}[t]
    \includegraphics[width=0.94\columnwidth]{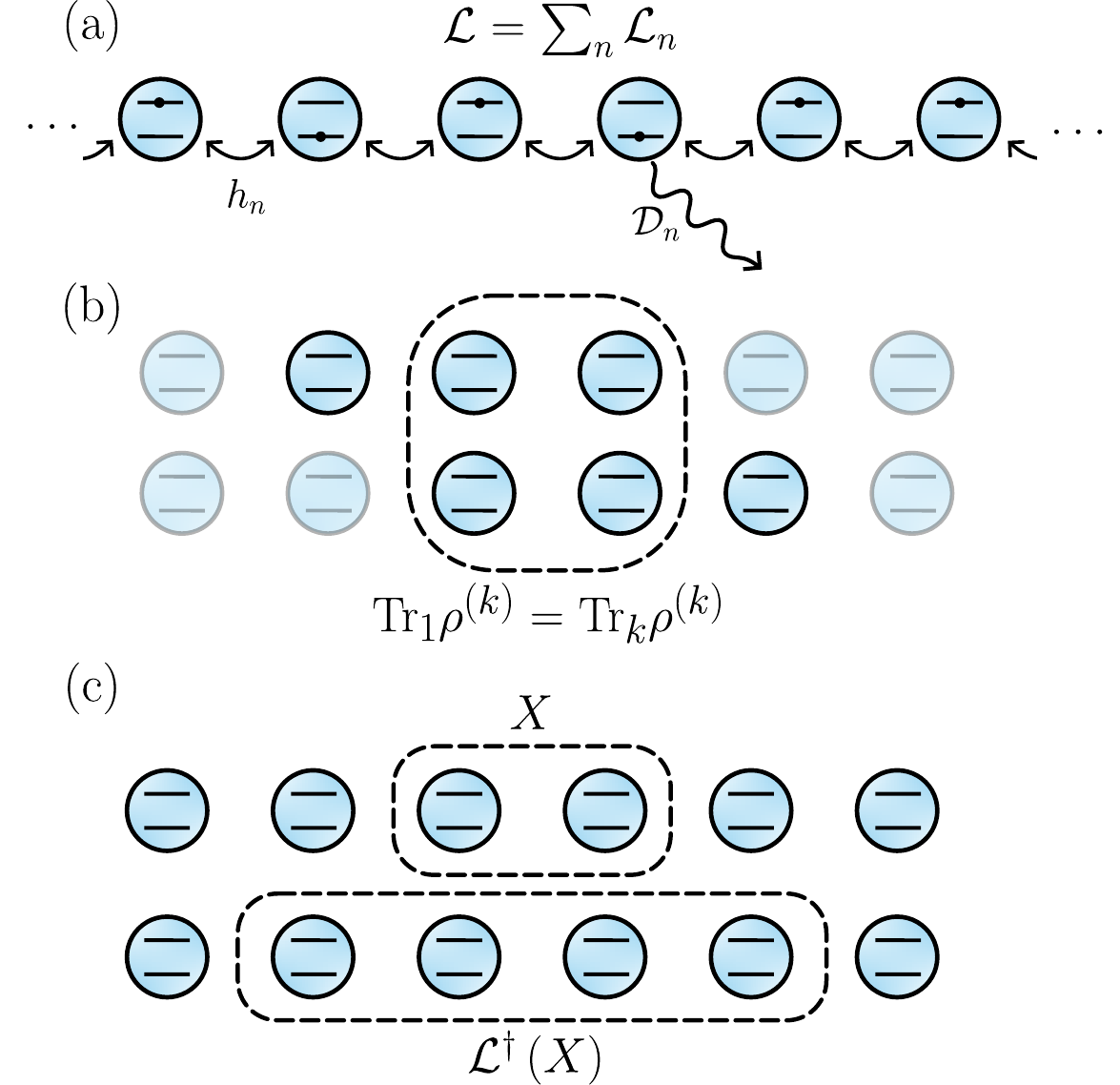}
    \caption{Diagrammatic depiction of the setup considered. (a) We target the steady state of an open system, here shown for the one-dimensional case. The dynamics of the system are generated by a Lindbladian which is a sum of local terms. (b) All reduced states of $k$ contiguous particles are described by the same density matrix, which has to satisfy the local translation invariance condition Eq.~(\ref{eq:lti}). (c) Due to the strict locality of the Lindbladian, the growth of the support of an operator $X$ when acted upon by the adjoint Lindbladian is bounded. For a Lindbladian with two-body terms, an operator with support on $k'$ sites extends to an operator of size $k'+2$ after the action of the adjoint Lindbladian.}
    \label{fig_1}
\end{figure}

\paragraph{One-dimensional translation-invariant systems.---}Consider the reduced density matrix $\rho^{\left( k \right)}$ of $k$ contiguous particles in a translation-invariant system of size $N$. A valid reduced state has to be positive, $\rho^{\left( k \right)} \geq 0$, and normalized, $\textrm{Tr} \rho^{\left( k \right)} = 1$. In addition, due to translation invariance (TI), all possible reduced states of such a subsystem are described by the same density matrix. Compatibility between all of them implies the constraint
\begin{equation}\label{eq:lti}
    \textrm{Tr}_1 \rho^{\left( k \right)} = \textrm{Tr}_k \rho^{\left( k \right)},
\end{equation} 
where $\textrm{Tr}_1$ ($\textrm{Tr}_k$) denotes the partial trace over the leftmost (rightmost) qudit of the subsystem (see Fig.~\ref{fig_1}(b)). Mathematically, the three previous conditions are satisfied by all reduced density matrices of one-dimensional translation-invariant states, though they are not sufficient to ensure extendability. Furthermore, we can combine them with the steady state condition~\eqref{eq:stst}, which can be rewritten as

\begin{equation} \label{eq:stationary_constraint}
    \textrm{Tr} \left(X \mathcal{L}\left(\rho_s\right) \right) = \textrm{Tr} \left(\mathcal{L}^\dagger \left(X\right) \rho_s \right) = 0,
\end{equation} for an arbitrary operator $X$ and where $\mathcal{L}^\dagger$ is the adjoint Lindbladian,

\begin{equation}
    \mathcal{L}^\dagger(X) = i \left[ H, X\right] + \sum_n L^\dagger_n X L_n - \frac{1}{2}\{L^\dagger_n L_n, X \}.
\end{equation} For a local Lindbladian, the action of the adjoint Lindbladian increases the support of an operator in proportion to the size of the boundary of its original support. In particular, as shown in Fig.~\ref{fig_1}(c), in one-dimensional systems with two-body Lindbladians, if an operator $X$ is supported on $k'$ sites, then  $\mathcal{L}^\dagger\left(X\right)$ is supported on at most $k'+2$ sites. Moreover, as long as $k \geq k'+2$, we have 

\begin{equation} \label{eq:stationary_constraint_2}
    \textrm{Tr} \left( \mathcal{L}^\dagger \left( X \right) \rho_s \right) =  \textrm{Tr} \left( \mathcal{L}^\dagger \left( X \right) \rho^{(k)} \right) = 0.
\end{equation} This means that the stationary constraint in Eq.~\eqref{eq:stationary_constraint} can be translated to the reduced density matrix $\rho^{(k)}$ for all the operators that have support on $k-2$ sites. 

Let us now consider the scenario in which we are interested in the expectation value of a local operator $O$ supported on at most $k$ sites on the steady state $\rho_s$ of a TI one-dimensional system. Conditions of the form~\eqref{eq:stationary_constraint_2}, together with~\eqref{eq:lti}, are necessary, but generally not sufficient, to specify the set of possible reduced states $\rho^{(k)}$ compatible with the global state being in the stationary state. As they involve only a positive semidefinite constraint and linear equalities, they can be characterized by a semidefinite program (SDP) \cite{vandenberghe1996}. In this framework, we can bound the expectation value of $O$ through the following optimization problem:

\begin{equation} \label{eq:sdp_rdm}
\begin{aligned}
\underset{\rho^{\left( k \right)}}{\textrm{max/min}}\quad & \textrm{Tr} \left( O \rho^{\left( k \right)}\right)\\
\textrm{s.t.} \quad & \rho^{\left( k \right)} \geq 0, \textrm{Tr} \rho^{\left( k \right)} = 1, \\
                    & \textrm{Tr}_1 \rho^{\left( k \right)} = \textrm{Tr}_k \rho^{\left( k \right)}, \\
                    & \textrm{Tr} \left(\mathcal{L}^\dagger \left(X_i \right) \rho^{\left( k \right)} \right) = 0. \\
\end{aligned}
\end{equation} Here, the operators $X_i$ define a basis for the space of operators with support on $k-2$ contiguous sites, indexed by $i$. These are precisely the set of operators such that the constraint in Eq.~\eqref{eq:stationary_constraint} can be imposed at the level of $\rho^{(k)}$. Note, however, that the previous SDP can also be posed for a set of operators $X_i$ that do not span the entire operator space of $k-2$ contiguous sites. The solution of the optimization problem~\eqref{eq:sdp_rdm} bounds the range of values the expectation value $\langle O \rangle$ can take on the steady state. We refer to this interval as the \textit{allowed region} or the \textit{bounds} of the expectation value. 

Let us comment on some of the properties of the optimization problem we have introduced: (i) this method is a relaxation of the problem of computing expectation values in the steady state. The relaxation would be exact if we worked with the density matrix of the entire system, $\rho^{(N)}$, and we imposed the stationary constraint for a complete basis $X_i$ of operators in the global Hilbert space, $\textrm{Tr} \left(\mathcal{L}^\dagger \left(X_i \right) \rho^{\left( N \right)} \right) = 0$. (ii) Increasing $k$ gives a sequence of increasingly tighter bounds --- the feasible set becomes strictly smaller as we make $k$ larger. This can be seen by checking that any reduced state $\rho^{(k)}$ in the feasible set for a given size $k$ would also be in the feasible set of any smaller size, provided we take the partial trace over the appropriate qudits. The converse is not true.  (iii) The optimization problem we have formulated is independent of the total system size $N$. This means that solutions of the previous SDP for a given size $k$ bound the result that we would obtain if we solved exactly the problem $\mathcal{L}\left(\rho_s\right)=0$ for a TI system of arbitrary size $N\geq k$. In particular, the derived bounds also apply to the thermodynamic limit. As such, our results can be understood as a numerical proof of the convergence of finite-size results to the thermodynamic limit. This is analogous to the behavior of the Anderson bound \cite{anderson1951, kull2024, eisert2023} for ground states of finite and infinite systems.

\paragraph{Non-translation-invariant systems.---}
So far, we have assumed translation invariance to formulate our relaxation for computing steady-state expectation values in open quantum systems. This assumption is commonly adopted in numerical approaches and is often well motivated, as approximate TI typically emerges in the bulk of large many-body systems. Here, we briefly explain why TI is required in our formulation and outline how the method can be generalized when it is absent.

In the SDP of \ref{eq:sdp_rdm}, TI is enforced through the constraint $\mathrm{Tr}_1 \rho^{(k)} = \mathrm{Tr}_k \rho^{(k)}$. Without it, the SDP would apply to all Lindbladians that coincide locally but differ arbitrarily outside the considered region of $k$ sites, potentially leading to drastically different steady states. This limitation can be overcome by introducing as variables the reduced density matrices of all $k$ contiguous subsystems, $\rho^{(1\dots k)}, \rho^{(2\dots k+1)}, \dots, \rho^{(N-k+1\dots N)}$, each constrained to be positive and normalized. Consistency between overlapping regions is enforced by
\begin{equation}
\mathrm{Tr}_i \rho^{(i\dots k+i-1)} = \mathrm{Tr}_{k+i} \rho^{(i+1\dots k+i)}, \quad i = 1,\dots,N-k.
\end{equation}
As in Eq.~\ref{eq:sdp_rdm}, stationarity conditions $\textrm{Tr} \left( \mathcal{L}^\dagger \left( X_i \right) \rho_s \right) = 0$ can be imposed for all operators $X_i$ such that all the terms in $\mathcal{L}^\dagger\left(X_i\right)$ appear in the available reduced density matrices. The resulting SDP has a computational cost that scales linearly with the system size $N$. Results based on this approach will be presented in upcoming work~\cite{frias2025prep}.

\begin{figure}[t]
    \includegraphics[width=\columnwidth]{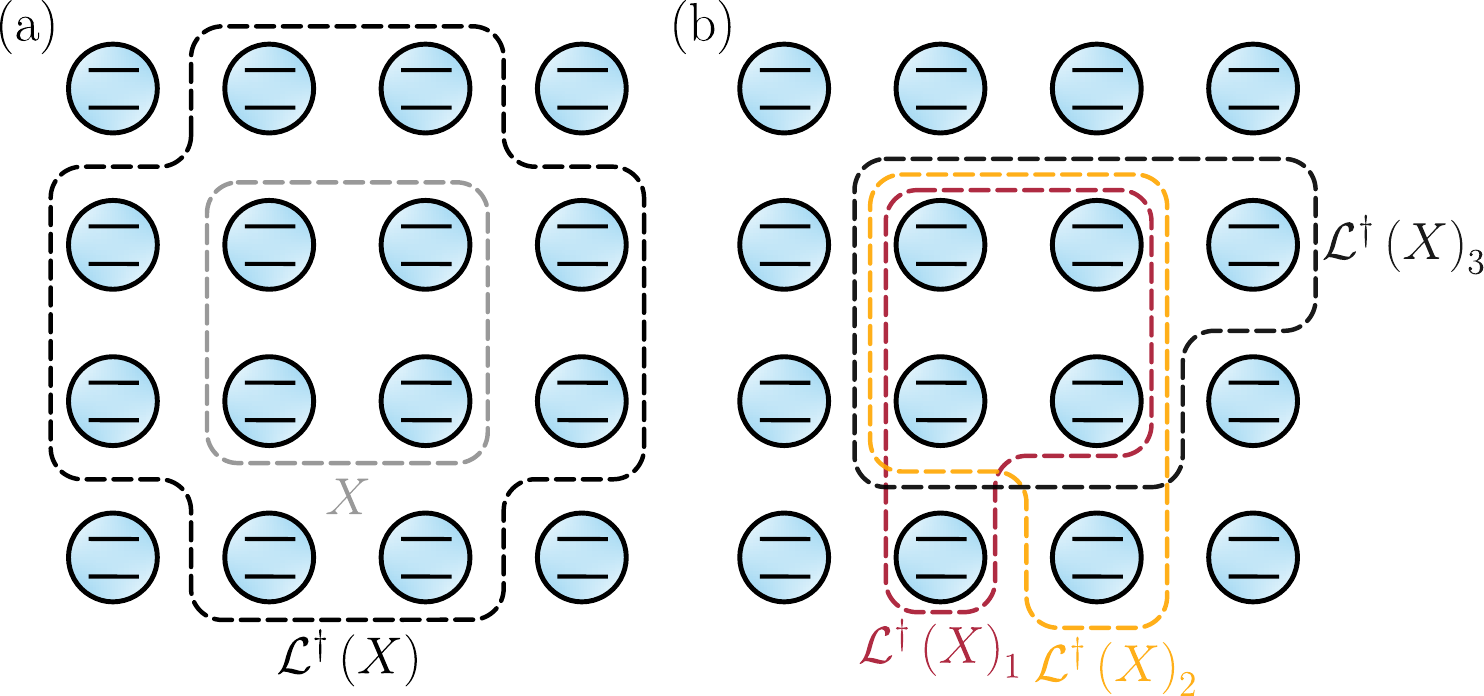}
    \caption{Diagram of the implementation of our method for two-dimensional open systems. (a) Consider an operator $X$ supported on a region of the lattice. The action of the adjoint Lindbladian on the operator, $\mathcal{L}^\dagger \left( X \right)$, is an operator supported on a larger region. The particular growth of the operator is determined by the connectivity of the Lindbladian. If it acts only on nearest-neighbours, the support of the operator grows by one unit in the directions in which the Lindbladian acts non-trivially. (b) The operator $\mathcal{L}^\dagger \left( X \right)$, however, can be written as a sum of terms, $\left( \mathcal{L}^\dagger \left( X \right)\right)_n$, where each term acts non-trivially on the initial support of $X$ and one extra qudit. If the support of the original operator $X$ is symmetric under rotation and reflection symmetries, the expectation value of some of the terms $\left( \mathcal{L}^\dagger \left( X \right) \right)_n$ can be related to others.}
    \label{fig_2}
\end{figure}

\begin{figure*}
    \centering
    \includegraphics[width=0.245\linewidth]{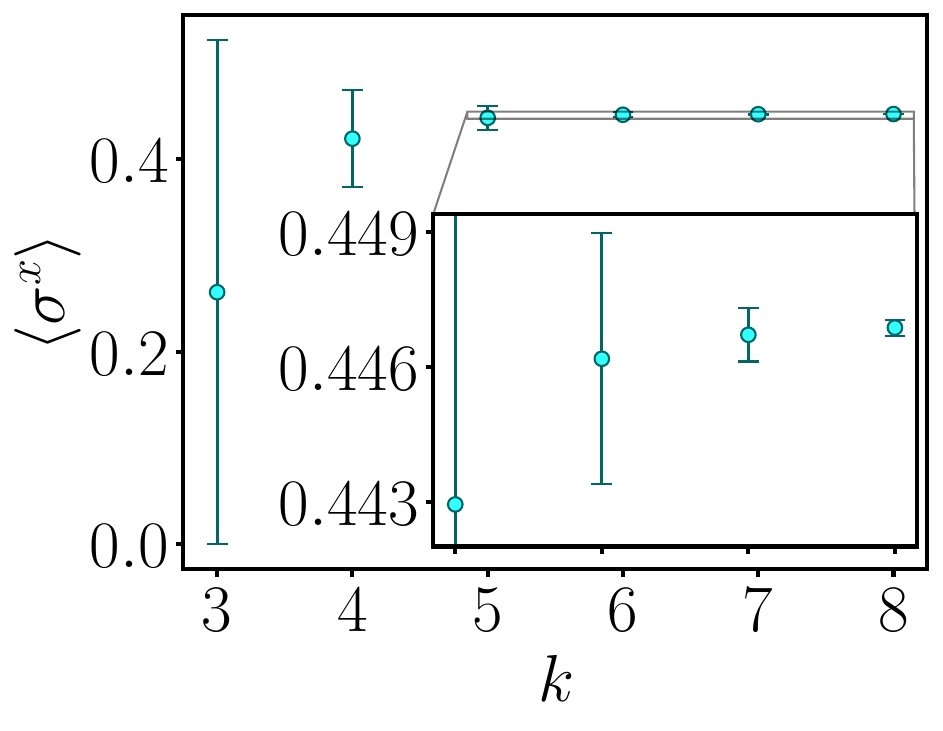}
    \includegraphics[width=0.245\linewidth]{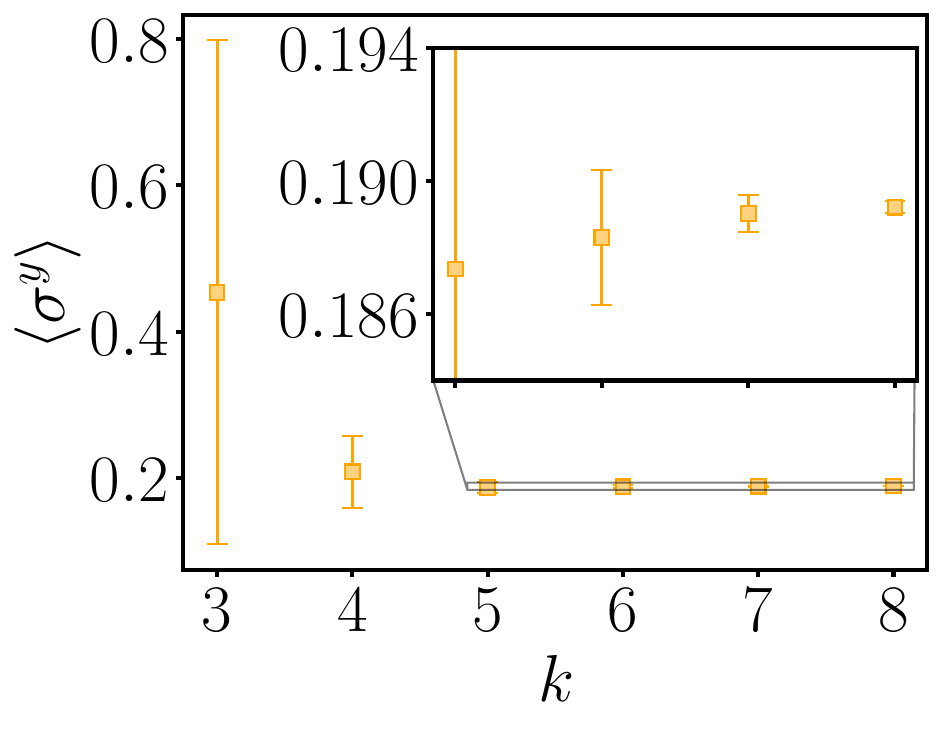}
    \includegraphics[width=0.255\linewidth]{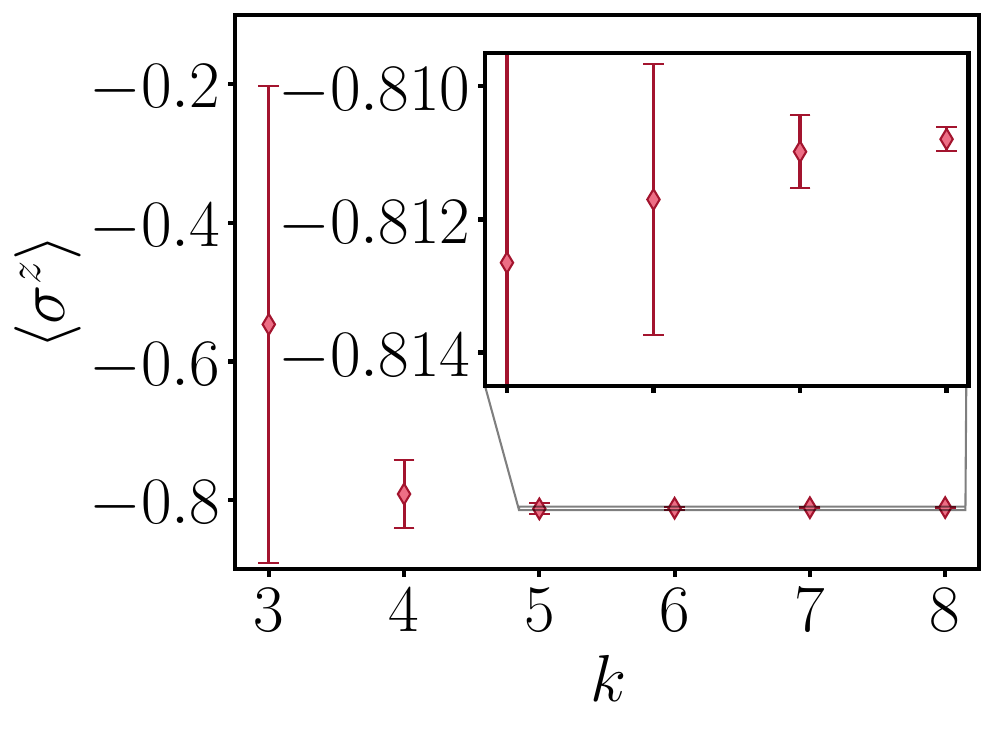}
    \includegraphics[width=0.235\linewidth]{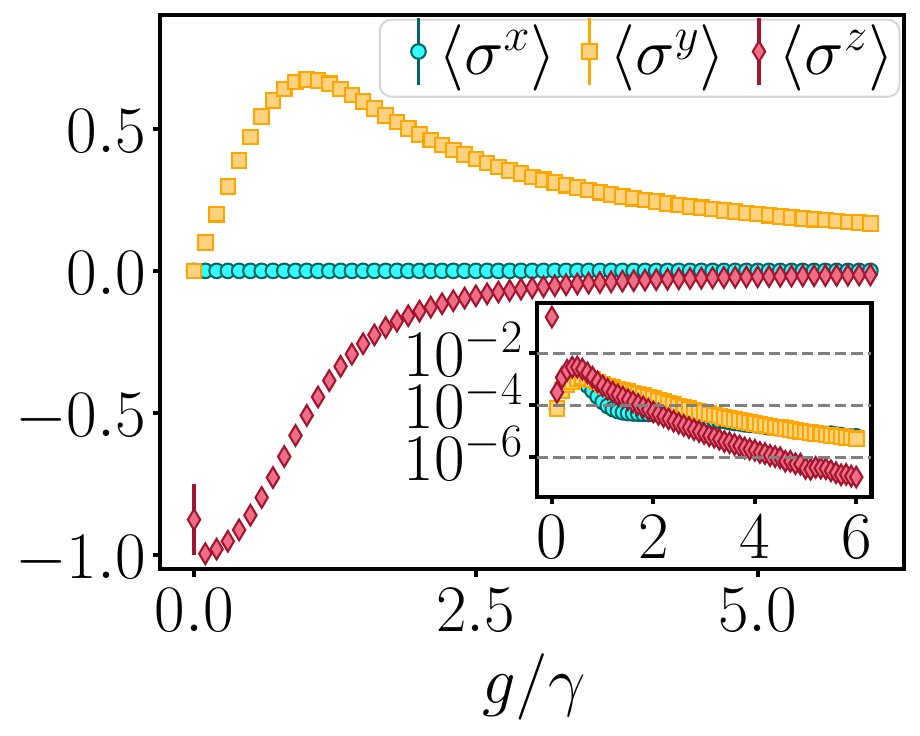}
    \caption{Bounds obtained with our method for the local expectation values of two different one-dimensional models. In all the main figures, markers indicate the center of the allowed region for each operator, while the error bars represent is full extent. (a)-(c) Bounds on the expectation value of (a) $\langle \sigma^x\rangle$, (b) $\langle \sigma^y\rangle$, and (c) $\langle \sigma^z\rangle$ for the one-dimensional dissipative Ising model  as a function of $k$, the size of the reduced density matrix $\rho^{(k)}$ in the SDP~\eqref{eq:sdp_rdm}. The insets show a zoom-in of the results for larger values of $k$, as the error bars become much smaller than the markers. (d) Results obtained with $k = 8$ for the one-dimensional short-range Dicke model as a function of $g/\gamma$. Except for $\langle \sigma^z \rangle$ at $g/\gamma = 0$, the error bars are smaller than the markers. For reference, the inset displays the size of the allowed region for the same expectation values.}
    \label{fig_3}
\end{figure*}

\paragraph{Higher-dimensional systems.---} The method extends straightforwardly to higher dimensions. For concreteness, we focus on a two-dimensional square lattice; other geometries can be treated analogously. Let $\Lambda$ be a connected region of the lattice and $\rho^{(\Lambda)}$ its reduced density matrix, which must satisfy
\begin{equation}
\rho^{(\Lambda)} \ge 0, \qquad \mathrm{Tr}\,\rho^{(\Lambda)} = 1 .
\end{equation} Spatial symmetries are incorporated by requiring invariance under the lattice symmetry group $G$. For each $g \in G$, with unitary representation $U_g$, this is enforced through the linear constraints
\begin{equation}
\mathrm{Tr}\!\left[\bigl(X_i - U_g X_i U_g^\dagger\bigr)\rho^{(\Lambda)}\right] = 0 ,
\end{equation} for a basis of operators $X_i$ supported on $\Lambda$ such that $U_g X_i U_g^\dagger$ is also supported on $\Lambda$. These constraints account for translation, rotation, and reflection symmetries of the square lattice, if present.

As in the one-dimensional case, symmetry constraints can be combined with the local stationarity condition
\begin{equation}
\mathrm{Tr}\!\left(\mathcal{L}^\dagger(X_i)\rho^{(\Lambda)}\right) = 0 ,
\end{equation} for all operators $X_i$ such that $\mathcal{L}^\dagger(X_i)$ has support contained in $\Lambda$, ensuring compatibility with a global steady state. In general, the growth of the support when acting with $\mathcal{L}^\dagger$ on an operator $X_i$ is proportional to the boundary of the support of $X_i$ (see Fig.~\ref{fig_2}(a)). As the cost of solving the SDP grows exponentially with the size of the region $\Lambda$, this can set a practical limitation on the size of the operators $X_i$ whose stationarity we can impose. A way to alleviate it, illustrated in Fig.~\ref{fig_2}, is to consider the structure of the operator $\mathcal{L}^\dagger \left(X_i\right)$. For a Lindbladian with two-body terms, we can write

\begin{equation}
    \mathcal{L}^\dagger \left(X_i\right) = \sum_n \left(\mathcal{L}^\dagger \left(X_i\right)\right)_n.
\end{equation} If $X_i$ has support on $k'$ sites, each of the $\left(\mathcal{L}^\dagger \left(X_i\right)\right)_n$ has support on $k'+1$ sites. The stationarity condition can then be imposed at the level of the reduced density matrices of $k'+1$ particles, provided we include all the necessary ones to evaluate all the terms. Furthermore, if the support of $X_i$ is symmetric under some of the lattice symmetries, we can reduce the number of density matrices we need to include in our SDP, as illustrated in Fig.~\ref{fig_2}(b).

\paragraph{Numerical experiments.---} To illustrate the algorithm's performance, we apply it to several open many-body spin models, where the competition between unitary and dissipative dynamics gives rise to steady states with non-trivial correlations.

\paragraph{Ising model.---} As a first benchmark, we consider a dissipative version of the Ising model in one dimension, with Hamiltonian $~{H = \sum_n J \sigma^{z}_n \sigma^{z}_{n+1} + g \sigma^{x}_n}$ and jump operators $~{L_n = \sqrt{\gamma} \sigma^{-}_n/2}$. To make a connection with previous results in the literature \cite{mortimer2025,hryniuk2024}, we choose $J = g = 1/2$ and $\gamma = 1$. The bounds on the single-qubit expectation values of the steady state obtained using our method are shown in Fig.~\ref{fig_3}(a)-(c) for different sizes of the reduced density matrix $\rho^{(k)}$ in the relaxation. For all the expectation values, the bounds show very fast convergence with $k$. Furthermore, the convergence is monotonic, as the size of the feasible set cannot increase with $k$. For the largest reduced density matrix we can simulate, $k = 8$, we obtain the following bounds on the local state, $ \langle \left( \sigma_x, \sigma_y, \sigma_z \right)\rangle = \left( 0.4469 \pm 0.0002, 0.1892 \pm 0.0002, -0.8108 \pm 0.0002\right)$. These bounds are rigorous and valid for systems of all sizes $N$ larger than $k$, even in the thermodynamic limit. 

\paragraph{One-dimensional short-range Dicke model.---} We study next a model of independent two-level systems, modeled with the Hamiltonian $H = g \sum_{n} \sigma^{x}_{n}$, coupled through the jump operators $L_n =  \gamma \left( \sigma^{-}_{n} + \sigma^{-}_{n+1} \right)$ which induce correlations in the system through dissipation. This model is inspired by the phenomenon of superradiance in the Dicke model \cite{dicke1954, gross1982}, where independent two-level systems coupled through a single collective jump operator, $L = \sum_n \sigma^{-}_n$, experience enhanced correlated decay in the limit of small dissipation. In the one-dimensional setting, at $g/\gamma=0$, the model has a two-fold degenerate steady-state space, spanned by $\ket{1}^{\otimes N}$ and $\sum_n (-1)^n\ket{1^{(n-1)}01^{(N-n)}}/\sqrt{N}$. This degeneracy can be a limitation for other methods in the regime of small $g/\gamma$ \cite{cui2015}, as non-positive operators appear in the kernel of the Lindbladian superoperator. Our method resolves the degeneracy, as can be seen in Fig.~\ref{fig_3}(d), from the errorbar at $g/\gamma = 0$, which stretches the range of expectation values any mixture of the two steady states can take. For $g/\gamma \neq 0$, our method gives accurate predictions for the entire phase diagram, predicting at least two decimals of all the expectation values in the single-qubit reduced density matrix. However, close to $g/\gamma \approx 1$, the errorbars grow in size, which we attribute to a build-up of correlations in the system.

\begin{figure}[t]
    \includegraphics[width=0.49\columnwidth]{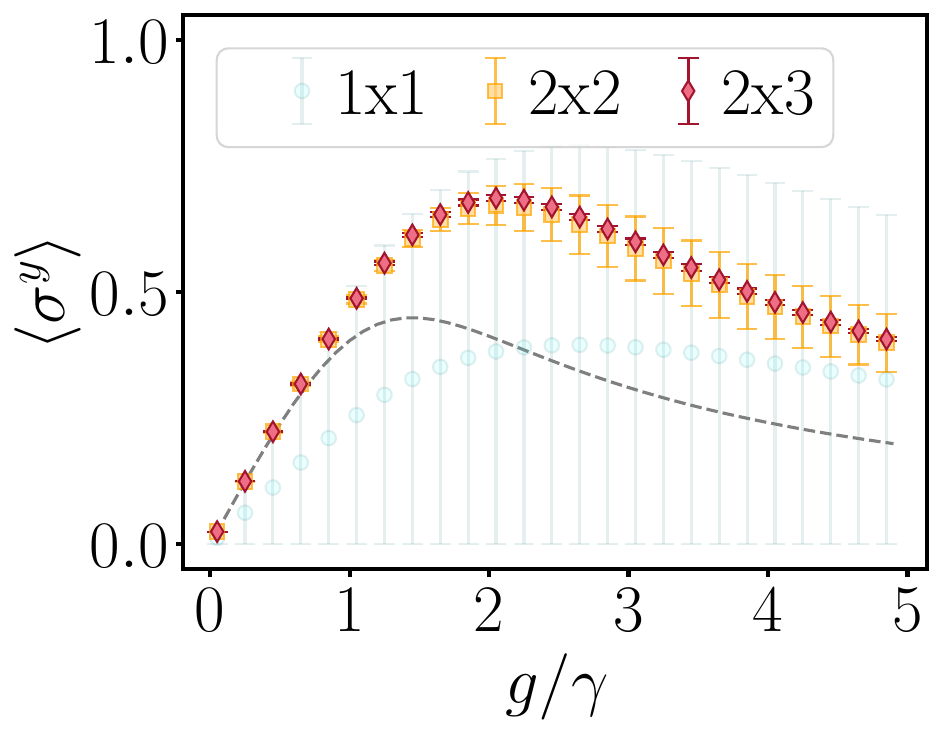}
    \includegraphics[width=0.49\columnwidth]{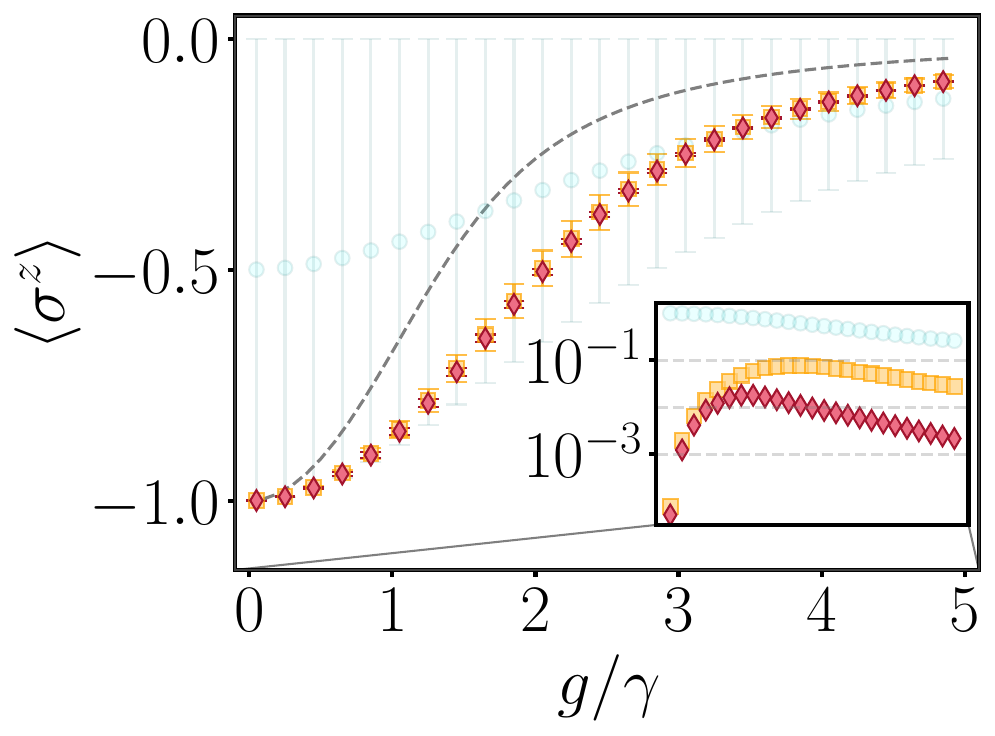}
    \caption{Bounds on the local expectation values (a) $\langle \sigma^y\rangle$ and (b) $\langle \sigma^z\rangle$ for the phase diagram of the steady state of the considered two-dimensional short-range Dicke model. Each error bar shows different sizes of the reduced density matrix used to impose the stationarity of the local expectation values. In both figures, the dashed gray line shows the result of a mean-field computation, whose validity is ruled out by our method. In the inset of (b), we show the size of the error bars of $\langle \sigma^z\rangle$ for the different regions we use as a function of $g/\gamma$. The horizontal lines show, as an indicator, an error bar size of $10^{-3}$, $10^{-2}$, and $10^{-1}$, respectively. We see that with the largest reduced density matrix, we can get results accurate to two decimal digits through practically the entire phase diagram.}
    \label{fig_4}
\end{figure}

\paragraph{Two-dimensional short-range Dicke model.---} We study last the behavior of the model in the previous section in a two-dimensional lattice, with Hamiltonian $H = \sum_{ij} \sigma^{x}_{ij}$, and jump operators in the two directions of the lattice, $~{L_{i,j}^{(h)} = \gamma \left( \sigma^{-}_{i,j} + \sigma^{-}_{i+1,j} \right)}$, and $~{L_{i,j}^{(v)} = \gamma \left( \sigma^{-}_{i,j} + \sigma^{-}_{i,j+1} \right)}$. The results obtained with our method are shown in Fig.~\ref{fig_4} and compared with those from a mean-field analysis. For a given square (or rectangular) cluster of particles, we can impose the stationarity condition Eq.~\eqref{eq:stationary_constraint} of all the operators supported on it, provided we add to the reduced density matrices some particles around to be able to read off the constraints from them (see Fig.~\ref{fig_2}). For clusters of sizes $1\times1$, $2\times2$, $2\times3$ this gives rise to reduced density matrices of size $2$, $5$ and $9$, respectively. For the largest cluster that we can simulate, we obtain bounds in the expectation value of single-qubit Pauli operators whose absolute size is below $0.018$ for $\langle \sigma_y \rangle$ and $0.013$ for $\langle \sigma_z \rangle$. Notice that $\langle \sigma_x \rangle = 0$ can be shown for the entire phase diagram of this model by symmetry arguments.

\paragraph{Discussion.---} By shifting the focus from the global many-body state to the reduced density matrices of local subsystems, we have introduced an algorithm that bounds the value a given observable can take in the steady state of a many-body open quantum system. When combined with the symmetries present in the problem, the bounds become very precise and apply to systems of arbitrary size or even the thermodynamic limit. Due to the rigorous nature of our algorithm, it can be used as well to benchmark other classical or quantum approaches to solve for steady states.

Looking ahead, it would be interesting to study the performance of the algorithm near phase transitions and to understand the connection between the convergence of the bounds and the spectral properties of the Lindbladian \cite{cubitt2015}. Furthermore, our algorithm could be combined with other relaxation approaches \cite{mortimer2025, araujo2025, lawrence2024} to further improve the results or study the dynamical relaxation towards the steady state.

\begin{acknowledgments}

We thank A. Leitherer, L. Mortimer, P. Mujal, and M. Pawlik, for collaboration in the early stages of the project, and M. C. Bañuls, D. Farina, and R. Trivedi for discussions. This project has received funding from the Government of Spain (Severo Ochoa CEX2019-000910-S, FUNQIP and Quantum in Spain), Fundació Cellex, Fundació Mir-Puig, Generalitat de Catalunya (CERCA program), the European Union (Quantera Veriqtas), the ERC AdG CERQUTE and the AXA Chair in Quantum Information Science.


\end{acknowledgments}

\bibliographystyle{apsrev4-1}
\bibliography{rdms_steady_states}

\end{document}